\documentclass[pra,aps,showpacs,twocolumn,]{revtex4}

\usepackage{amssymb}
\usepackage{graphicx}

\begin{document}

\title{On the role of hyperfine-interactions-mediated Zeeman effect in the condensation temperature shift of trapped atomic Bose-Einstein condensates}

\author{Fabio Briscese$^{1,2}$, Sergei Sergeenkov$^{1}$, Marcela Grether$^3$ and Manuel de Llano$^4$}
\affiliation{$^{1}$ Departamento de F\'isica, CCEN, Universidade
Federal da Para\'iba, Cidade Universit\'aria, 58051-970 Jo\~ao
Pessoa, PB, Brazil.\\
$^{2}$ Istituto Nazionale di Alta Matematica Francesco Severi,
Gruppo Nazionale di Fisica Matematica, Citt\`a Universitaria, P.le
A. Moro 5, 00185 Rome, EU.\\
$^3$ Facultad de Ciencias, Universidad Nacional Aut\'onoma de
M\'exico, 04510 M\'exico, DF, M\'exico\\
$^4$ Instituto de Investigaciones en Materiales, Universidad
Nacional Aut\'onoma de M\'exico, A. P. 70-360, 04510 M\'exico, DF,
M\'exico}


\begin{abstract}
We discuss the effect of interatomic interactions on the
condensation temperature $T_c$ of a laboratory atomic
Bose-Einstein condensate under the influence of an external
trapping magnetic field. We predict that accounting for hyperfine
interactions mediated Zeeman term in the mean-field approximation
produces, in the case of the $403 \, G$ Feshbach resonance in the
$|F,m_F> = |1,1>$ hyperfine state of a $^{39}K$ condensate, with
$F$ the total spin of the atom, an experimentally observed (and
not yet explained) shift in the condensation temperature $\Delta
T_{c}/T_{c}^{0}=b^{*}_0+b^{*}_1 (a/\lambda_{T}) + b^{*}_2
(a/\lambda_{T})^2$ with $b^{*}_0 \simeq 0.0002$, $b^{*}_1 \simeq
-3.4$ and $b^{*}_2 \simeq 47$, where $a$ is the s-wave scattering
length, and $\lambda_T$ is the thermal wavelength at $T_{c}^{0}$.
Generic expressions for the coefficients $b^*_0$, $b^*_1$ and
$b^*_2$ are also obtained, which can be used to predict the
temperature shift for other Feshbach resonances of $^{39}K$ or
other atomic condensates.
\end{abstract}

\maketitle

The study of Bose-Einstein condensates (BECs) is a important
current subject of modern physics (see
Refs\cite{dalfovo,parkinswalls,burnett} for a review). Atomic BECs
are produced in the laboratory in laser-cooled,
magnetically-trapped ultra-cold bosonic clouds of different atomic
species, e.g., $_{37}^{87}Rb$ \cite{Ander},$\ _{3}^{7}Li$
\cite{Bradley}, $ _{11}^{23}Na$ \cite{Davis}, $_{1}^{1}H$
\cite{Fried}, $_{37}^{87}Rb$ \cite {Cornish}, $_{2}^{4}He$
\cite{Pereira}, $_{19}^{41}K$ \cite{Mondugno}, $ _{55}^{133}Cs$
\cite{Grimm}, $_{70}^{174}{Yb}$ \cite{Takasu03} and $
_{24}^{52}Cr$ \cite{Griesmaier}. Also, BECs of photons are
nowadays under investigation \cite{becs of photons}. Moreover,
BECs are commonly applied in cosmology and astrophysics \cite{becs
cosmology} and in fact have been shown also to constrain quantum
gravity models \cite{briscese}.

In the context of atomic BECs, inter-particle interactions play a
fundamental role, since they are necessary to drive the atomic
cloud to thermal equilibrium, so they must carefully be taken into
account when studying the properties of the condensate. For
instance, interatomic interactions change the condensation
temperature $T_c$ of a BEC, as  was pointed out first by Lee and
Yang \cite{a1,a2} (see also
\cite{a3,a4,a5,a6,a7,a8,a9,a10,a11,a12} for more recent works).

The first studies  of  interactions effects were focused on
uniform BECs. Here, interactions are absent in the mean field (MF)
approximation (see \cite{a7,a10,a11,a12} for reviews) but they
produce a shift in the condensation temperature with respect to
the ideal noninteracting case, which is due to beyond-MF effects
related to quantum correlations between bosons near the critical
point. This effect has been finally quantified in \cite{a7,a8} as
$\Delta T_{c}/T_{c}^{0}\simeq \, 1.8 \, (a/\lambda_{T})$, where
$\Delta T_{c}\equiv T_{c}-T_{c}^{0}$ with $\,T_{c}$ the critical
temperature of the gas of interacting bosons, $T_{c}^{0}$ the
condensation temperature in the ideal non-interacting case,
$\lambda_{}\equiv \sqrt{2 \pi \hbar^2/(m_a k_B T^0_c)}$ the
thermal wavelength at temperature $T^0_c$, $m_a$ the atom mass,
and $a$  the s-wave scattering length used to parameterize
inter-particle interactions \cite{dalfovo,parkinswalls,burnett}.

It should be noted that laboratory condensates are not uniform
BECs since they are produced in atomic clouds confined in magnetic
traps, but they can be described in terms of harmonically-trapped
BECs consisting of a system of $N$ bosons trapped in an external
spherically symmetric harmonic potential $V = m_a\omega^2 x^2/2$,
with $\omega$ the frequency of the trap. For trapped BECs,
interactions affect the condensation temperature even in the MF
approximation, and the shift in $T_c$ in terms of the s-wave
scattering length is given by

\begin{equation}
\Delta T_{c}/T_{c}^{0}\simeq b_1 (a/\lambda_{T}) + b_2
(a/\lambda_{T})^2. \label{deltaTsuNONTUNIFORM1}
\end{equation}
with $b_1 \simeq -3.426$ \cite{dalfovo} and $b_2 \simeq 18.8$
\cite{briscese EPJB}, implying that  $\Delta T_{c}$ is negative
for repulsive interactions \footnote{We recall that  the
condensation temperature of a non-interacting harmonically trapped
condensate  is $k_B T^0_c = \hbar \omega
\left(N/\zeta(3)\right)^{1/3}$
 \cite{dalfovo}}.

High precision measurements \cite{condensatePRL} of the
condensation temperature of $^{39}K$ in the range of parameters
$N\simeq (2-8) \times 10^5$, $\omega \simeq (75-85) Hz$, $10^{-3}
< a/\lambda_{T} < 6 \times 10^{-2}$ and $T_c \simeq (180-330) nK$
have detected second-order effects in $\Delta T_{c}/T_{c}^{0}$
fitted by the following expression $\Delta T_{c}/T_{c}^{0} \approx
b_1^{exp} (a/\lambda_{T}) + b_2^{exp} (a/\lambda_{T})^2$ with
$b_1^{exp} \simeq -3.5 \pm 0.3$ and $b_2^{exp} \simeq 46 \pm 5$.
This result has been achieved exploiting the $403G$ Feshbach
resonance in the $|F,m_F> = |1,1>$ hyperfine state of a $^{39}K$
condensate, where $F$ is the total (nuclear + electronic) spin of
the atom. Therefore, if the theoretically predicted linear
contribution $b_1$ was found to practically coincide with the
observed value, its nonlinear (quadratic) counterpart $b_2$ turned
out to be in strong disagreement with experimental data.

It should be mentioned that there have been some efforts to
theoretically estimate the correct value of  $b_2$ in the MF
approximation, for instance considering anharmonic and even
temperature-dependent traps \cite{castellanos},  which however
have been not successful. Therefore one could expect that a more
realistic prediction of the experimental value of $b_2^{exp}$
should take into account beyond-MF effects.

The goal of this paper is to show that, taking into account the
Zeeman effect and using the MF approximation, it is quite possible
to predict the experimentally observed value of $b_2 \simeq 47$
for the $403G$ resonance of the hyperfine $|F,m_F> = |1,1,>$ state
of $^{39}K$ as measured in \cite{condensatePRL}, with no need to
appeal to any beyond-MF effects. We also discuss the generality of
this result and the possibility of predicting the condensation
temperature shift for different resonances of $^{39}K$ and for
different atomic condensates.

Recall that in the MF Hartree-Fock approximation (assuming the
semiclassical condition $k_B T \gg \hbar \omega$), bosons are
treated as a noninteracting gas that experiences a MF interaction
potential $\propto g n(x)$, where $g = (4 \pi \hbar^2 a/m_a)$
\cite{dalfovo,parkinswalls,burnett} and $n(x)$ is the local
density of bosons at the point $x$, so that the Hartree-Fock
hamiltonian is\cite{dalfovo,parkinswalls,burnett}

\begin{equation}\label{HF Hamiltonian}
H_{HF} =\frac{P^2}{2m_a} + V(x) + 2 g n(x)
\end{equation}

It is important to remind that experimentally the s-wave
scattering length parameter $a$ is tuned via the
Feshbach-resonance technique based on Zeeman splitting of bosonic
atom levels in applied magnetic field. It means that $g$ in
Eq.(\ref{HF Hamiltonian}) is actually {\it always}
field-dependent, since $g \propto a=a(B)$.  More explicitly,
according to the interpretation of the Feshbach resonance
\cite{derrico,williams}

\begin{equation}\label{a B}
a(B)=a_0\left (1-\frac{\Delta}{B-B_0}\right)
 \label{feshbach}
\end{equation}
where $a_0$ is the so-called background value of the length, $B_0$
is the resonance peak field, and $\Delta$ the width of the
resonance.

Thus, in order to properly address the problem of condensation
temperature shifting (which is always observed under application
of a non-zero magnetic field $B$), one has to add to Eq. (\ref{HF
Hamiltonian}) a missing hyperfine interactions mediated Zeeman
contribution $H_Z=-\mu(x)B$ where $\mu(x)$ is the local magnetic
moment of a Bose atom in the trap. Since an applied magnetic field
affects the condensate, this moment depends on the local density
$n(x)$ as follows, $\mu(x)=\mu_an(x)V_m$ where $\mu_a=g_sS\mu_N$
is the magnetic moment of a particular atom with $S$ the nuclear
spin, $g_s$ the gyromagnetic coefficient, and $\mu_N=e\hbar/2m_p$
the nuclear magneton ($m_p$ being the proton mass). Here $V_m=4\pi
a_m^3$ is a characteristic volume of the condensate affected by
hyperfine interactions between atoms with $a_m$ being a magnetic
analog of the scattering length $a$. According to the
spectroscopic data \cite{derrico}, there are singlet ($a_S$) and
triplet ($a_T$) scattering lengths.

It can be easily verified that accounting for the Zeeman
contribution in Eq.(\ref{HF Hamiltonian}) will result in a simple
renormalization of the interaction constant $g(B)$ (which depends
on applied magnetic field via the s-wave length $a(B)$ given by
Eq.(\ref{a B})) as follows

\begin{equation}
g^{*}(B)=g(B)-\frac{1}{2}\mu_aBV_m
 \label{feshbach2}
\end{equation}
and the corresponding scattering length
\begin{equation}
a^{*}(B)=a(B)-\alpha B
 \label{feshbach3}
\end{equation}
with $\alpha=\mu_am_aV_m/8\pi \hbar^2$.

Now by inverting Eq.(\ref{a B}) and expanding the resulting $B(a)$
dependence into the Taylor series, one obtains

\begin{equation}\label{a B 2}
a^{*}=a-\alpha B_0\left(1+\frac{\Delta}{B_0}\right)-\alpha \Delta
\left[\frac{a}{a_0}+ \left(\frac{a}{a_0}\right)^2+...\right]
 \label{feshbach3a}
\end{equation}
for an explicit form of the renormalized (due to Zeeman splitting)
scattering length $a^{*}(B)$. Now, to find the changes of the
amplitudes $b_1$ and $b_2$ in the presence of the Zeeman effect,
we simply replace the original (Zeeman-free) scattering lengths
$a$ in Eq.(\ref{deltaTsuNONTUNIFORM1}) with their renormalized
counterparts $a^{*}$ (given by Eq. (\ref{a B 2})) which will
result in the following expression for the temperature shift

\begin{equation}\label{deltaTsuNONTUNIFORM 2}
\frac{\Delta T_{c}}{T_{c}^{0}}=b_1
\left(\frac{a^{*}}{\lambda_{T}}\right) + b_2
\left(\frac{a^{*}}{\lambda_{T}}\right)^2
 \label{feshbach4}
\end{equation}
Now, by using Eq.(\ref{a B 2}) we can rewrite
Eq.(\ref{deltaTsuNONTUNIFORM 2}) in terms of the original
scattering lengths $a$ and renormalized amplitudes $b^{*}_i$ as
follows

\begin{equation} \label{deltaTsuNONTUNIFORM 3}
\frac{\Delta T_{c}}{T_{c}^{0}}=b_0^{*}+b_1^{*}
\left(\frac{a}{\lambda_{T}}\right) + b^{*}_2
\left(\frac{a}{\lambda_{T}}\right)^2
 \label{feshbach5}
\end{equation}
where the new amplitudes (due to the Zeeman contribution) are
governed by the following expressions

\begin{equation} \label{b star 1}
b^{*}_0=-\left(\frac{\xi}{\lambda_{T}}\right)b_1+\left(\frac{\xi}{\lambda_{T}}\right)^2b_2,
\end{equation}

\begin{equation}\label{b star 2}
b^{*}_1=(1-\gamma)b_1-2(1-\gamma)\left(\frac{\xi}{\lambda_{T}}\right)b_2,
\end{equation}

and

\begin{equation}\label{b star 3}
b^{*}_2=\left[(1-\gamma)^2
+2\gamma\left(\frac{\xi}{a_0}\right)\right
]b_2-\gamma\left(\frac{\lambda_{T}}{a_0}\right)b_1
\end{equation}
where $\gamma=\alpha \Delta/a_0$, and $\xi=\alpha (B_0+\Delta)$.

Note that accounting for Zeeman effect resulted in the appearance
of a constant amplitude $b^{*}_0$. As we shall demonstrate below,
this contribution is very small and does not affect the
experimentally observed temperature shift.

To fix the model parameters, we proceed as follows. First of all,
we quite reasonably assume that Zeeman effect does not change the
linear contribution by putting $b^{*}_1=b_1$. Secondly, to find
the absolute change of the second amplitude due to Zeeman term, we
assume that $b^{*}_2=c b_2$ where $c$ is a constant (amplifying
factor). In view of Eqs.(\ref{b star 1}-\ref{b star 3}), the above
two assumptions bring about the following analytical expression
for the seeking amplifying parameter

\begin{equation}\label{c}
c=1+\frac{2g_sS\mu_Nm_aB_0a_m^3}{\hbar^2a_0}=1+S\left(\frac{B_0}{B_m}
\right)
\end{equation}
which is the main result of this paper. To obtain the second form
of the above expression, we have introduced a characteristic
magnetic field $B_m$ related to hyperfine interactions. More
precisely, $B_m=\Phi_0/\sigma_m$ where $\Phi_0=h/2e=2\times
10^{-15}Wb$ is the flux quantum and the projected area $\sigma_m$
is given by $\sigma_m=\pi g_s(m_a/m_p)a_m^3/a_0$. Note that, as
expected, in the absence of Zeeman effect (when $\mu_a=0$), we
have $c=1$ and thus $b^{*}_2=b_2$.

Let us consider the particular case of the $403G$ resonance of the
hyperfine $|F,m_F> = |1,1,>$ state of $^{39}K$. According to
\cite{derrico}, the relevant parameters needed to create and
measure magnetically trapped bosons for this atom are as follows:
$S=3/2$, $g_s=1/2$, $m_a=39 m_p$, $B_0=403G$, $a_0=-29 r_B$ (where
$r_B=0.053nm$ is the Bohr radius), and $a_S=138 r_B$ (for the
singlet magnetic scattering length). According to our Eq.(12), the
above set of parameters produces $c\simeq 2.5$ which readily leads
to the following estimate of the quadratic amplitude contribution
due to the Zeeman effect, $b^{*}_2=2.5 b_2\simeq 47$ in a good
agreement with observations\cite{condensatePRL}. It is interesting
to point out that since the nuclear spin of $^{39}K$ is $S=3/2$,
the obtained value $c\simeq 2.5$ for the amplifying factor means
that we have practically a complete match between the Feshbach
resonance field $B_0$ and the hyperfine interaction related field
$B_m$, namely $B_m\simeq B_0$.

To check self-consistency of our calculations, we also estimated
the value of the constant amplitude $b^{*}_0$ (which is equal to
zero in the absence of the Zeeman effect). The result is
$b^{*}_0\simeq 0.0002$ which is, as expected, a rather negligible
contribution, even though it is not zero.

In conclusion,  we have shown that accounting for
hyperfine-interactions-induced Zeeman term in the mean-field
approximation produces,  for the $403 \, G$ Feshbach resonance in
the $|F,m_F> = |1,1>$ hyperfine state of a $^{39}K$ condensate, an
experimentally observed shift of the condensation temperature
$T_c$  given by Eq. (\ref{deltaTsuNONTUNIFORM 3}) with $b^{*}_0
\simeq 0.0002$, $b^{*}_1 \simeq -3.4$ and $b^{*}_2 \simeq 47$.

It would be interesting to put the predicted universal relation
(\ref{c}) to further experimental test in order to find out
whether it can also explain the values of $b^*_2$ for other
resonances of $^{39}K$ as well as for other atomic condensates by
repeating the measurements of the second-order interactions
effects performed in Ref. \cite{condensatePRL}.

This work has been financially supported by the Brazilian agencies
CNPq and CAPES. MdeLl thanks PAPIIT-UNAM for grant IN-100314 and
MG for grant IN-116914.

\end{document}